\documentclass{aastex}
\usepackage{emulateapj5}

\begin{document}

\newcommand{\hi}{$h^{-1}$~}
\newcommand{\kms}{~km~s$^{-1}$}
\newcommand{\logh}{$+5\log h$}

\title{Spectroscopic Confirmation of Multiple Red Galaxy-Galaxy Mergers
  in MS~1054--03 ($z=0.83$)$^{1,2,3}$}

\author{Kim-Vy H. Tran$^4$,  Pieter van Dokkum$^5$, Marijn Franx$^6$,
Garth D. Illingworth$^7$, Daniel D. Kelson$^8$, \& 
Natascha M. F\"orster Schreiber$^9$}

\footnotetext[1]{Based on observations with the NASA/ESA Hubble Space
Telescope, obtained at the Space Telescope Science Institute, which
is operated by the Association of Universities for Research in
Astronomy, Inc., under NASA contract NAS 5-26555.}
\footnotetext[2]{Based on observations obtained at the W. M. Keck 
Observatory, which is operated jointly by Caltech and the University
of California.} 
\footnotetext[3]{Based on observations obtained at the the European
  Southern Observatory, Paranal, Chile.}
\footnotetext[4]{Institute for Astronomy, ETH Z\"urich, CH-8093
Z\"urich, Switzerland, vy@phys.ethz.ch} 
\footnotetext[5]{Department of Astronomy, Yale University, New Haven,
CT 06520-8101}
\footnotetext[6]{Leiden Observatory, P.O. Box 9513, 2300 RA Leiden,
The Netherlands}
\footnotetext[7]{University of California Observatories/Lick
Observatory, University of California, Santa Cruz, CA 95064}
\footnotetext[8]{Observatories of the Carnegie Institution of
Washington, 813 Santa Barbara Street, Pasadena, CA, 91101}
\footnotetext[9]{Max-Planck-Institut f\"ur Extraterrestrische Physik,
Giessenbachstrasse, 85748 Garching bei M\"unchen, Germany}
        
\setcounter{footnote}{9}

\begin{abstract}
  
  We present follow-up spectroscopy of the galaxy cluster MS~1054--03
  ($z=0.83$) confirming that at least six of the nine merging galaxy
  pairs identified by \citet{vandokkum:99} are indeed bound systems:
  they have projected separations of $R_s<10$\hi~kpc and relative
  line-of sight velocities of $\Delta v<165$\kms.  For the remaining
  three pairs, we were unable to obtain redshifts of both constituent
  galaxies.  To identify a more objective sample of merging systems, we
  select bound red galaxy pairs ($R_s\leq30$\hi~kpc, $\Delta
  v\leq300$\kms) from our sample of 121 confirmed cluster members:
  galaxies in bound red pairs make up $15.7\pm3.6$\% of the cluster
  population.  The $(B-K_s)$ color-magnitude diagram shows that the
  pair galaxies are as red as the E/S0 members and have a homogeneous
  stellar population.  The red pair galaxies span a large range in
  luminosity and internal velocity dispersion to include some of the
  brightest, most massive members ($L>L^{\ast}$,
  $\sigma_{1D}>200$\kms); these bound galaxy pairs must evolve into
  E/S0 members by $z\sim0.7$.  These results combined with MS1054's
  high merger fraction and reservoir of likely future mergers indicates
  that most, if not all, of its early-type members evolved from (passive)
  galaxy-galaxy mergers at $z\lesssim1$.

\end{abstract}


\section{Introduction}

Whether galaxy-galaxy merging plays a prominent role in the formation
of early-type galaxies in clusters remains an open question.  The
expected merger rate in rich, virialized clusters is low because the
relative velocities of members are too high
\citep[$e.g.$][]{makino:97}.  Observational studies are consistent with
this idea as the fraction of galaxies in merging systems is low in
clusters at $z\lesssim0.3$ \citep[$e.g.$][]{dressler:80,couch:98}.  In
this context, the discovery by \citet[hereafter vD99]{vandokkum:99}
that 17\% of the galaxies in the rich cluster MS~1054--03 ($z=0.83$)
are in merging systems was a surprise.  vD99 argue that because MS1054
has not yet virialized, the high merger rate in this cluster is
``permitted.''

The main uncertainty in the vD99 study is that the merging systems were
identified by eye, and identifying bound galaxy pairs in a more
objective manner is needed to isolate a representative sample of
merging systems.  Such a sample provides better constraints on the
end-product of the mergers and how galaxy-galaxy merging changes the
global cluster population.  For example, significant amounts of merging
between red members will alter the cluster's stellar mass function.
The most robust approach for selecting these ``future'' mergers is via
a combination of projected galaxy separation and relative line-of-sight
velocity, but it does require a great deal of spectroscopy.

Motivated by these issues, we revisit the issue of galaxy-galaxy
merging in MS1054 using an expanded sample of spectroscopically
confirmed cluster members.  In this paper, we use $(\Omega_M,
\Omega_{\Lambda}, H_0)$ of $(0.3, 0.7, 70)$; at $z=0.83$, this gives a
physical scale of $7.6$~kpc per arcsecond.

\section{Summary of Observations}

\subsection{Optical and Near-Infrared Imaging}

MS1054 was part of a large program carried out by our group to study
galaxy clusters at intermediate redshift.  As part of our observing
campaign, we obtained wide-field ($R_{BCG}\sim1$\hi~Mpc), high
resolution imaging of MS1054 with HST/WFPC2 in F606W and F814W.
Details of the image reduction and morphological classification are in
\citet{vandokkum:00}.

In addition, deep wide-field ground-based imaging of MS1054 in $UBV$
and $J_sHK_s$ was taken on the VLT with FORS1 and ISAAC, respectively.
Details of the image reduction and photometry can be found in
\citet{forster:05}.  To summarize, all of the images were convolved to
a common spatial resolution of FWHM$=0.69''$.  Objects then were
detected using SExtracter v2.2.2 \citep{bertin:96} and colors measured
using custom isophotal apertures.

\subsection{Keck/LRIS Spectroscopy}

To determine cluster membership, we observed MS1054 using LRIS
\citep{oke:95} on the Keck Telescope from February 1997 to June 2001.
During 6 observing runs, 21 multi-slit masks were used to measure
redshifts and internal velocity dispersions for $I$-selected targets;
one mask (June 2001) specifically targeted classified mergers.  Details
on how the spectra were reduced, redshifts determined, and spectral
indices measured ([OII]$\lambda3727$, H$\delta$, \& H$\gamma$) are in
\citet{tran:99,tran:03b}.  The typical error associated with the
redshifts is $\delta v\sim30$\kms.

Considering only galaxies that fall on the HST/WFPC2 mosaic, we have
121 confirmed cluster members that span a range in luminosity
($-22.9<M_{Be}-5\log h< -18.7$); for these galaxies, we have measured
structural parameters, colors, and morphologies.  Our analysis also
includes measured internal velocity dispersions ($\sigma_{1D}$) for 26
members \citep{kelson:01,wuyts:04}.

\section{Visually Identified Galaxy Mergers}

We have obtained follow-up spectroscopy for the original 13 mergers
reported by vD99 to determine if they are truly bound systems.  To
clarify, WFPC2 imaging showed that these mergers were made of 9 close
pairs and 4 (single) morphologically irregular galaxies; we refer to
\citet[Fig. 8]{postman:05} for higher resolution imaging from the Advanced
Camera for Surveys (ACS) of the same systems.  We focus here on the 9
close pairs.

Redshifts of the constituent galaxies for 6 of the 9 close pairs
confirms that these are bound systems: they have projected separations
of $R_s<10$\hi~kpc and relative velocities of $\Delta v<165$\kms.  To
confirm that their relative velocities are not due to chance
projections between members, we compare their $\Delta v$ to that of a
random (normalized) cluster sample where the latter is generated by
drawing $10^4$ galaxy pairs from the full redshift sample of cluster
members (Fig.~\ref{combo}, left).  A Kolmogorov-Smirnov test shows the
two distributions differ at the $>99.7$\% level.

For the remaining 3 close pairs, we were unable to obtain redshifts for
the companion galaxies: one companion was not targeted, and the spectra
of the other 2 were too faint to measure a redshift.  However, we
stress that all 6 of the close pairs for which we have measured
relative velocities confirm that they are merging systems.  Although we
were unable to obtain relative velocities for the remaining 3 close
pairs from vD99, it is very likely that they too are bound systems.

\section{Objectively Identified Galaxy Pairs}

\subsection{Counting Close Galaxy Pairs}

As a quantitative measure of the excess of close pairs in MS1054, we
follow \citet{vandokkum:00b} and examine the distribution of
galaxy-galaxy separations in the WFPC2 photometric catalog.  For each
galaxy, we count the number of objects $N$ that lie at a given
distance $R_s$ and then weight by the surface area to obtain the
surface density. By summing over all galaxies for a given
galaxy-galaxy separation, we then have the relative galaxy density
$\delta$ as a function of $R_s$.  For simplicity, we bin the sample
into rings defined by $R_{max}$ and $R_{min}$ where $R_s$ is the
average of these two values.  Thus for a given ring, we have:

\begin{equation} 
\delta = \frac{\sum_{i}^{j} (N^i)}{\pi (R_{max}^2-R_{min}^2)}
\end{equation}

\noindent where $j$ is the total number of galaxies.  If the galaxies
are distributed uniformly, the distribution of $\delta$ as a function
of $R_s$ will be flat.

Because we are primarily interested in the red cluster members, we
consider only the 142 galaxies from the WFPC2 catalog with
$(R_{606}-I_{814})>1.7$ and $I_{814}<23$ mags; for reference, the
brightest cluster galaxy (BCG) has $(R_{606}-I_{814})=2.13$ and
$I_{814}=19.5$ mags.  Our color and magnitude selection are effective
at isolating red cluster members: we have measured redshifts for 91 of
these red galaxies and 89 (98\%) are confirmed cluster members.

Figure~\ref{combo} (middle) shows the relative galaxy density $\delta$
as a function of increasing $R_s$ for these red galaxies.  Since the
high galaxy density in the cluster core can increase $\delta$ for small
separations, we exclude the inner $R_{BCG}<200$\hi kpc region.  The
$\delta$ distribution is essentially flat except for a strong peak at
$R_s<10$\hi kpc.  

To further investigate the issue of projected versus bound pairs, we
consider the 5 red pairs with $R_s<10$\hi~kpc and 7 red pairs with
$10\leq R_s<20$\hi~kpc at $R_{BCG}>200$\hi~kpc.  From the average
relative density of pairs with $R_s>50$\hi~kpc, we estimate that 4/5
and 3/7 are bound pairs with the remainder being chance projections.  We 
test this result with our spectroscopy. We find that all 5 red galaxy pairs
with $R_s<10$\hi~kpc are bound; note that all 5 also were identified by
vD99 as mergers.  Spectroscopy for 5 of the 7 red pairs with $10\leq
R_s<20$\hi~kpc shows that only 1 pair is bound.

\subsection{Defining a Sample of Bound Red Galaxy Pairs}

As demonstrated above, spectroscopy is critical for determining
whether photometric pairs ($R_s>10$\hi~kpc) are bound systems.  With
this in mind, we now define a spectroscopically confirmed sample of
bound red galaxy pairs.  First we take the 142 red galaxies and make
each one into a pair with every other red galaxy.  We then consider
only the red pairs that are made of confirmed cluster members and
measure their relative velocities and projected separations
(Fig.~\ref{combo}, right).

Guided by studies on merging pairs \citep{patton:00,lin:04}, we select
bound pairs as members with projected separations of
$R_s\leq30$\hi~kpc and relative velocities of $\Delta v\leq300$\kms:
there are 10 such systems (Fig.~\ref{tnails}).  Thus the fraction of
red merging galaxies in MS1054 is $15.7\pm3.6$\% (19/121).  With this
definition, a single galaxy can contribute to two separate bound
pairs; there is one such triplet system in our pair sample.  Note that
this differs from vD99 where merging pairs were defined to be a single
system.  We also verified that our spectroscopic sample is not biased
towards red galaxies with small projected separations: to
$I_{814}=22.5$ mags, the spectroscopic sample is $79\pm5$\% complete
compared to $69\pm14$\% complete for red galaxies with a neighbor
within $R_s\leq30$\hi~kpc.

As with all observational studies on galaxy pairs, we cannot say with
absolute certainty that our 10 bound red pairs will merge into a single
system by $z\sim0$.  However, detailed simulations of major mergers
find that the constituent galaxies merge in $0.5-2$ Gyr
\citep[e.g.][]{barnes:88,mihos:94b,springel:00,jonsson:05}. In
addition, \citet{patton:00} showed that the typical merging timescale
for galaxy pairs with $R_s\leq30$ kpc and $\Delta v\leq500$\kms~is 0.5
Gyr: given our bound pairs have even smaller projected separations and
relative velocities ($R_s\leq25$\hi~kpc, $\Delta v\leq300$\kms), it is
very likely they will merge by $z\sim0.7$.

\section{Discussion}

The physical properties of the galaxies making up MS1054's bound red
pairs indicate that the pairs can evolve into massive E/S0 members by
$z\sim0.7$.  The color-magnitude diagram (Fig.~\ref{cmd}) shows that 1)
the pair galaxies span the range in luminosity and include the
brightest cluster members; 2) they are as red (old) as the E/S0
population; and 3) their scatter in $(B-K_s)$ color is only $\sim1.5$
times that of the E/S0's.  Their relatively small scatter compared to,
$e.g.$ MS1054's late-types, indicates that the pair galaxies have
homogeneous stellar populations; assuming these systems evolve
passively, their fractional age differences will decrease and their
scatter reduce accordingly.  Their co-added spectrum confirms that the
pair galaxies are dominated by old stellar populations
(Fig.~\ref{spectrum}).  The pair galaxies also include very massive
members: 9 have internal velocity dispersions $>200$\kms~
\citep{wuyts:04}.

Can the number of E/S0 members formed via late-time merging between red
members significantly alter MS1054's stellar mass function?  From their
study covering a wide range in environment, \citet{bell:04} claim that
the stellar mass function for red galaxies increases by a factor of two
between $z\sim1$ and now.  If true, adopting the simplest scenario
implies that every present-day cluster early-type undergoes a merger at
$z\lesssim1$.  Assuming a typical merger timescale of 0.5 Gyr and
$\Delta t_{z=1}=7.7$ Gyr, this requires that at a given time, only
$\sim6.5$\% of the galaxies need be in a merging system.  The merger
fraction in MS1054 is more than twice this value ($16$\%), and MS1054
has a rather astonishing merger rate of $\sim20$ Gyr$^{-1}$.

\begin{figure}
\epsscale{0.9}
\plotone{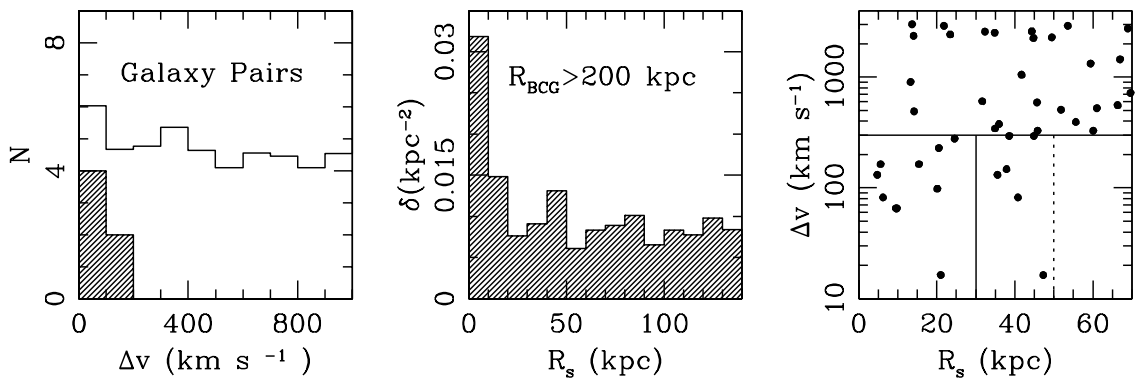}
\caption{{\it Left:} The relative line-of-sight velocity ($\Delta v$)
  of six merging galaxy pairs from vD99 (hatched) compared to the
  normalized distribution of $10^4$ galaxy pairs randomly drawn from
  the full cluster sample; all six merging pairs have projected
  separations of $R_s<10$\hi~kpc.  A K-S test confirms the two
  distributions differ at the $>99.7$\% level, thus the merging pairs
  are not due to chance projections between members and are bound
  systems.  {\it Middle:} The relative galaxy density ($\delta$) as a
  function of increasing galaxy-galaxy separation ($R_s$) in MS1054.
  Due to the high galaxy densities in the cluster core, we exclude the
  inner $R_{BCG}<200$\hi kpc region.  MS1054 has a strong peak at
  $R_s<10$\hi kpc; this excess of pairs at small separations
  independently confirms that MS1054 has an unusually large number of
  close red pairs. {\it Right:} Relative velocity ($\Delta v$) as a
  function of projected separation ($R_s$) for cluster pairs where
  redshifts of both galaxies have been measured.  We consider a bound
  pair to be a system composed of two galaxies with $\Delta
  v\leq300$\kms and $R_s<30$\hi kpc (solid lines): there are 10 such
  pairs.  MS1054 also has six red galaxy pairs with $30<R_s<50$\hi kpc
  (dotted line) and $\Delta v\leq300$\kms; these are likely to be
  future mergers.
\label{combo}}
\end{figure}

\begin{figure}
\plotone{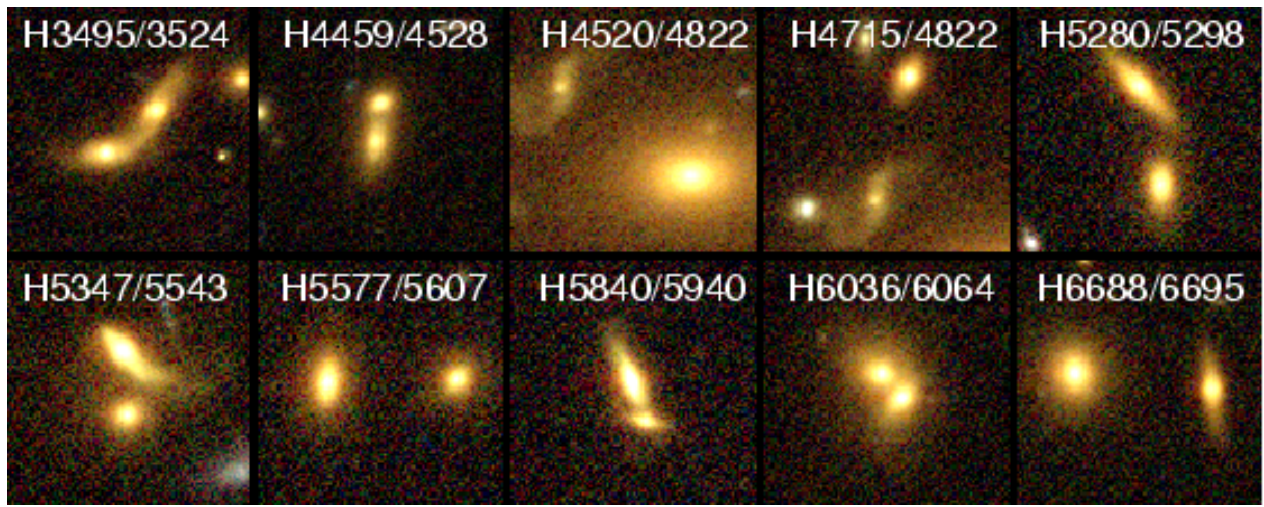}
\caption{{\bf [Higher resolution version available at 
http://www.exp-astro.phys.ethz.ch/tran/outgoing/ms1054mgrs.ps.gz]}
HST/ACS images ($5''\times5''$) of the 10 objectively
  selected, spectroscopically confirmed bound red galaxy pairs
  ($R_s\leq30$\hi~kpc, $\Delta v\leq300$\kms).  The five red pairs
  that were classified as mergers by vD99 are H3495/3524 (\#1340),
  H4459/4528 (\#1692), H5347/5543 (\#1760), H5840/5940 (\#1583), and
  H6036/6064 (\#1163).
\label{tnails}}
\end{figure}

\begin{figure}
\epsscale{0.9}
\plotone{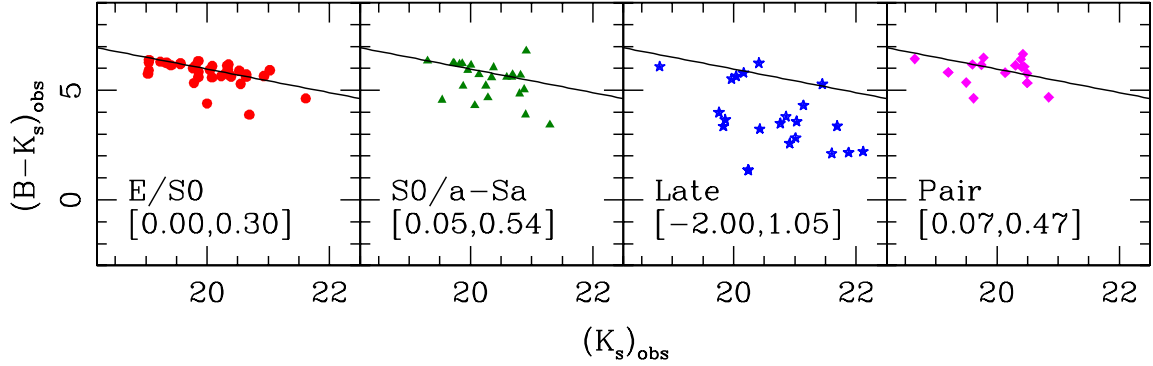}
\caption{ Color-magnitude (CM) diagram binned by
  galaxy type.  The CM relation shown in each panel is determined by
  fitting an iterative least-squares to the E/S0 members (solid line).
  Galaxies in bound red pairs span the range in luminosity and include
  the brightest cluster members.  We also list the deviation from the
  CM relation where we have measured the median offset ($C_{MED}$) and
  mean absolute deviation in color ($\sigma_{MAD}$; ``scatter'').  The
  pair galaxies are as red as the E/S0 members and their scatter in
  color is relatively small, especially compared to the late-type
  members.
\label{cmd}}
\end{figure}

\begin{figure}
\epsscale{0.4}
\plotone{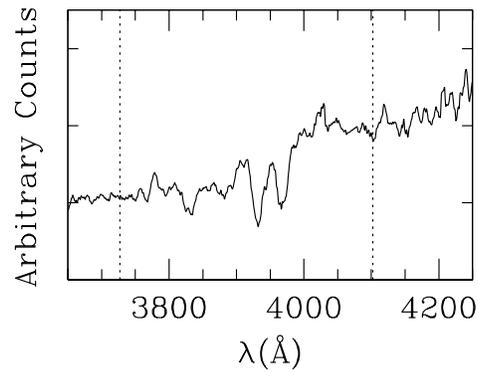}
\caption{Smoothed co-added spectra of cluster members in bound
  red pairs; the vertical dotted lines denote [OII]$\lambda3727$ and
  H$\delta$.  All spectra are normalized by their flux at
  4000\AA~before being co-added.  The strong break at 4000\AA~is
  characteristic of a stellar population dominated by old (G \& K)
  stars.
\label{spectrum}}
\end{figure}

Although MS1054's current merger rate is certainly high, the key
question is how long can it be sustained?  To increase the stellar mass
function of red galaxies by the quoted factor of two, MS1054 would need
to retain a merger fraction of $\sim16$\% over the next $\sim3$ Gyr
(assuming $\Delta t_{MGR}=0.5$ Gyr) and have all of these mergers
evolve into early-type members.  MS1054 does have a reservoir of possible
``future'' mergers: there are six red galaxy pairs with $30<R_s<50$\hi
kpc and $\Delta v\leq300$\kms~(Fig.~\ref{combo}, right) that can
provide at least one more generation of merging pairs.

Accretion of galaxy groups also can be a source of future mergers: the
group environment is conducive to galaxy-galaxy merging, and
Press-Schechter theory \citep{press:74} predicts MS1054 will more than
double its mass by $z\sim0$ (M.  Santos, private communication) via
accretion of, $e.g.$ groups \citep{gonzalez:05}.  MS1054's high merger
rate is very likely linked to its ongoing accretion of galaxy groups
\citep{hoekstra:00}; if so, this implies that an elevated merger rate
is associated with this special phase in cluster evolution.  MS1054's
merger rate will drop as it virializes \citep{makino:97}, a result
consistent with the low merger fractions found in lower redshift
clusters \citep[$e.g.$][]{couch:98}.

Thus in the case of MS1054, it is quite feasible that all of the
early-type members evolved from (passive) galaxy-galaxy mergers at
$z\lesssim1$, assuming the cluster does not virialize immediately after
the epoch of observation.

\acknowledgments
  
The authors thank D. Fabricant for help with the visual
classifications, D. Magee for help with the cluster redshift survey,
and M. Postman for use of the reduced ACS imaging.  K. Tran
acknowledges support from the Swiss National Science Foundation; she
also thanks P.  Norberg and M. Santos for helpful discussions.
Additional support from HST-GO09803.05-A and NAG5-7697 also are
acknowledged.  Finally, the authors extend special thanks to those of
Hawaiian ancestry on whose sacred mountain we are privileged to be
guests.

\bibliographystyle{/Users/vy/aastex/apj}
\bibliography{/Users/vy/aastex/tran}

\clearpage

\end{document}